\documentstyle[12pt]{article}

\catcode`\@=11
\long\def\@makefntext#1{
\protect\noindent \hbox to 3.2pt {\hskip-.9pt  
$^{{\ninerm\@thefnmark}}$\hfil}#1\hfill}		

\def\@makefnmark{\hbox to 0pt{$^{\@thefnmark}$\hss}}  
	
\def\ps@myheadings{\let\@mkboth\@gobbletwo
\def\@oddhead{\hbox{}
\rightmark\hfil\ninerm\thepage}   
\def\@oddfoot{}\def\@evenhead{\ninerm\thepage\hfil
\leftmark\hbox{}}\def\@evenfoot{}
\def\sectionmark##1{}\def\subsectionmark##1{}}

\renewcommand{\thefootnote}{\fnsymbol{footnote}}

\newcounter{sectionc}\newcounter{subsectionc}\newcounter{subsubsectionc}
\renewcommand{\section}[1] {\vspace*{0.6cm}\addtocounter{sectionc}{1} 
\setcounter{subsectionc}{0}\setcounter{subsubsectionc}{0}\noindent 
	{\normalsize\bf\thesectionc. #1}\par\vspace*{0.4cm}}
\renewcommand{\subsection}[1] {\vspace*{0.6cm}\addtocounter{subsectionc}{1} 
	\setcounter{subsubsectionc}{0}\noindent 
	{\normalsize\it\thesectionc.\thesubsectionc. #1}\par\vspace*{0.4cm}}
\renewcommand{\subsubsection}[1]
{\vspace*{0.6cm}\addtocounter{subsubsectionc}{1}
\noindent {\normalsize\rm\thesectionc.\thesubsectionc.\thesubsubsectionc. 
	#1}\par\vspace*{0.4cm}}

\newcounter{appendixc}
\newcounter{subappendixc}[appendixc]
\newcounter{subsubappendixc}[subappendixc]

\renewcommand{\appendix}[1] {\vspace*{0.6cm}
        \refstepcounter{appendixc}
        \setcounter{figure}{0}
        \setcounter{table}{0}
        \setcounter{equation}{0}
        \renewcommand{\thefigure}{\Alph{appendixc}.\arabic{figure}}
        \renewcommand{\thetable}{\Alph{appendixc}.\arabic{table}}
        \renewcommand{\theappendixc}{\Alph{appendixc}}
        \renewcommand{\theequation}{\Alph{appendixc}.\arabic{equation}}
        \noindent{\bf Appendix \theappendixc #1}\par\vspace*{0.4cm}}

\def\abstracts#1{{
\centering{\begin{minipage}{12.2truecm}\footnotesize\baselineskip=12pt\noindent
	\centerline{\footnotesize ABSTRACT}\vspace*{0.3cm}
	\parindent=0pt #1
	\end{minipage}}\par}} 

\renewenvironment{thebibliography}[1]
	{\begin{list}{\arabic{enumi}.}
	{\usecounter{enumi}\setlength{\parsep}{0pt}
\setlength{\leftmargin 1.25cm}{\rightmargin 0pt}
	 \setlength{\itemsep}{0pt} \settowidth
	{\labelwidth}{#1.}\sloppy}}{\end{list}}

\newcounter{itemlistc}
\newcounter{romanlistc}
\newcounter{alphlistc}
\newcounter{arabiclistc}

\newcommand{\fcaption}[1]{
        \refstepcounter{figure}
        \setbox\@tempboxa = \hbox{\footnotesize Fig.~\thefigure. #1}
        \ifdim \wd\@tempboxa > 6in
           {\begin{center}
        \parbox{6in}{\footnotesize\baselineskip=12pt Fig.~\thefigure. #1}
            \end{center}}
        \else
             {\begin{center}
             {\footnotesize Fig.~\thefigure. #1}
              \end{center}}
        \fi}

\newcommand{\tcaption}[1]{
        \refstepcounter{table}
        \setbox\@tempboxa = \hbox{\footnotesize Table~\thetable. #1}
        \ifdim \wd\@tempboxa > 6in
           {\begin{center}
        \parbox{6in}{\footnotesize\baselineskip=12pt Table~\thetable. #1}
            \end{center}}
        \else
             {\begin{center}
             {\footnotesize Table~\thetable. #1}
              \end{center}}
        \fi}

\def\@citex[#1]#2{\if@filesw\immediate\write\@auxout
	{\string\citation{#2}}\fi
\def\@citea{}\@cite{\@for\@citeb:=#2\do
	{\@citea\def\@citea{,}\@ifundefined
	{b@\@citeb}{{\bf ?}\@warning
	{Citation `\@citeb' on page \thepage \space undefined}}
	{\csname b@\@citeb\endcsname}}}{#1}}

\newif\if@cghi
\def\cite{\@cghitrue\@ifnextchar [{\@tempswatrue
	\@citex}{\@tempswafalse\@citex[]}}
\def\citelow{\@cghifalse\@ifnextchar [{\@tempswatrue
	\@citex}{\@tempswafalse\@citex[]}}
\def\@cite#1#2{{$\null^{#1}$\if@tempswa\typeout
	{IJCGA warning: optional citation argument 
	ignored: `#2'} \fi}}

 1
 1
 1

\font\ninerm=cmr9

\def\beq{\begin{eqnarray}}
\def\eeq{\end{eqnarray}}
\def\ee{\varepsilon}
\def\mprp{\mbox{\tiny $\bot$}}
\def\mprl{\mbox{\tiny $\|$}}

\def\pp{p_{\mbox{\tiny $\|$}}}
\def\zt{X_{\mbox{\tiny $\bot$}}}

\def\Pl{{\cal P}^\lambda}

\def\M{{\cal M}}

\def\lm{\lambda}
\def\half{\frac{1}{2}}
\newcommand{\eq}[1]{(\ref{#1})}
\newcommand{\prp}[1]{#1_{\mbox{\tiny $\bot$}}}
\newcommand{\prl}[1]{#1_{\mbox{\tiny $\|$}}}

\def\HHi{H\!\!\left(\frac{4 m^2}{\prl{(k')^2}}\right)}
\def\HHii{H\!\!\left(\frac{4 m^2}{\prl{(k'')^2}}\right)}
\def\ggg{\gamma \rightarrow \gamma \gamma}
\def\ggnunu{\gamma \gamma\to \nu \bar\nu } 
\def\frb{\gamma_{\mprl} \rightarrow \gamma_{\mprl} \gamma_{\mprp}}
\def\alw{\gamma_{\mprl} \rightarrow \gamma_{\mprp} \gamma_{\mprp}}
\def\zer{\gamma_{\mprl} \rightarrow \gamma_{\mprl} \gamma_{\mprl}}
\def\ff{\Lambda}
\def\tff{\widetilde \Lambda}
\def\1{1 \to 1 \, 2}
\def\2{1 \to 2 \, 2}

\begin{document}

\begin{flushright}
{\normalsize Yaroslavl State University\\
             Preprint YARU-HE-98/05\\
             hep-ph/9808246} \\[10mm]
\end{flushright}

\vspace*{0.9cm}
\centerline{\normalsize\bf THE TRANSITIONS {\large $\gamma \gamma\to \nu 
\bar\nu $} 
AND {\large $\gamma \to \gamma \gamma$}}
\baselineskip=16pt
\centerline{\normalsize\bf IN A STRONG MAGNETIC FIELD}
\vspace*{0.6cm}
\centerline{\footnotesize 
M.V.~CHISTYAKOV, A.V.~KUZNETSOV and N.V.~MIKHEEV}
\baselineskip=13pt
\centerline{\footnotesize\it 
Division of Theoretical Physics, Department of Physics,}
\baselineskip=12pt
\centerline{\footnotesize\it 
Yaroslavl State University, Sovietskaya 14,}
\baselineskip=12pt
\centerline{\footnotesize\it 
150000 Yaroslavl, Russian Federation}
\baselineskip=12pt
\centerline{\footnotesize E-mail: 
mch@uniyar.ac.ru, avkuzn@uniyar.ac.ru,} 
\baselineskip=12pt
\centerline{\footnotesize mikheev@yars.free.net}

\vspace*{0.9cm}
\abstracts{
The photon-neutrino process 
$\gamma \gamma\to \nu \bar\nu $ and photon splitting 
are considered in a strong magnetic field. 
The partial polarization amplitudes are calculated within the standard model 
in the limit of a strong field. The amplitudes do not depend on the 
field strength in this limit.
Using the vector parts of the amplitudes, the process of the photon 
splitting $\ggg$ is investigated both below and above the pair 
creation threshold. 
The splitting probabilities are calculated taking account of the photon 
dispersion and large radiative corrections near the resonance. 
}
 
\normalsize\baselineskip=15pt
\setcounter{footnote}{0}
\renewcommand{\thefootnote}{\alph{footnote}}

\vspace{20mm}

\begin{center}
{\it Talk presented at the 
Ringberg Euroconference \\ ``New Trends in Neutrino Physics'', 
Ringberg Castle, \\Tegernsee, Germany, 24 - 29 May 1998}
\end{center}

\newpage

\section{
Neutrino -- Two-Photon Vertex in Vacuum}

The electroweak process of the transition of two gammas to the
neutrino -- antineutrino pair is described by two Feynman diagrams with 
charged fermions in the loop and with the photon interchange. 



\def\markphotonatomur{\begin{picture}(2,2)(0,0)
                             \put(2,1){\oval(2,2)[tl]}
                             \put(0,1){\oval(2,2)[br]}
                     \end{picture}
                    }
\def\markphotonatomdr{\begin{picture}(2,2)(0,0)
                             \put(1,0){\oval(2,2)[bl]}
                             \put(1,-2){\oval(2,2)[tr]}
                     \end{picture}
                    }
\def\photonurhalf{\begin{picture}(30,30)(0,0)
                     \multiput(0,0)(2,2){5}{\markphotonatomur}
                  \end{picture}
                 }
\def\photondrhalf{\begin{picture}(30,30)(0,0)
                     \multiput(0,0)(2,-2){5}{\markphotonatomdr}
                  \end{picture}
                 }

\def\photonatomright{\begin{picture}(3,1.5)(0,0)
                                \put(0,-0.75){\tencircw \symbol{2}}
                                \put(1.5,-0.75){\tencircw \symbol{1}}
                                \put(1.5,0.75){\tencircw \symbol{3}}
                                \put(3,0.75){\tencircw \symbol{0}}
                      \end{picture}
                     }
\def\photonright{\begin{picture}(30,1.5)(0,0)
                     \multiput(0,0)(3,0){10}{\photonatomright}
                  \end{picture}
                 }
\def\photonrighthalf{\begin{picture}(30,1.5)(0,0)
                         \multiput(0,0)(3,0){5}{\photonatomright}
                      \end{picture}
                     }

\begin{minipage}[t]{100mm}

\unitlength=1.00mm
\special{em:linewidth 0.4pt}
\linethickness{0.4pt}

\vspace*{35mm}

\begin{picture}(100.00,7.00)(-20,10)
\put(30.00,38.00){\circle{16.00}}
\put(37.0,38.00){\circle*{2.3}}
\put(24.0,34.00){\circle*{1.0}}
\put(24.0,42.00){\circle*{1.0}}

\put(37.0,18.00){\makebox(0,0)[cc]{Fig. 1}}
\put(7.00,50.00){\makebox(0,0)[cc]{\large $\gamma(k_1)$}}
\put(7.00,26.00){\makebox(0,0)[cc]{\large $\gamma(k_2)$}}

\put(14.0,24.00){\photonurhalf}
\put(14.0,52.00){\photondrhalf}

\put(52.00,48.00){\makebox(0,0)[cc]{$\nu_i(p_1)$}}
\put(52.00,28.00){\makebox(0,0)[cc]{$\bar \nu_i(p_2)$}}

\put(37.0,38.50){\line(3,1){16.00}}
\put(37.0,38.50){\vector(3,1){10.00}}
\put(37.0,37.50){\line(3,-1){16.00}}
\put(37.0,37.50){\vector(3,-1){10.00}}

\put(76.00,38.00){\makebox(0,0)[cc]
{$+ \, (\gamma_1 \leftrightarrow \gamma_2)$}}
\end{picture}

\end{minipage}

\noindent
Here the bold circle corresponds to the weak effective interaction 
carried by both the $W$ and $Z$ bosons. 
The main interest to this process exists in astrophysics where it could be 
an additional channel of neutrino creation in a stellar thermal bath.

The most general 
amplitude could be written in the following form
\beq
{\cal M} = {\alpha \over \pi} \, {G_F \over \sqrt{2}}\,
\left [\bar\nu_i (p_1) \,
T_{\alpha \beta \mu \nu} \,
\nu_i (- p_2) \right ]\,
f^{\alpha \beta}_1 f^{\mu \nu}_2, 
\label{Mgen}
\eeq

\noindent
where $i$ is the neutrino flavor, $i = e, \mu, \tau$, and the tensors 
$f^{\alpha \beta} = k^\alpha \ee^\beta - k^\beta \ee^\alpha$ are the photon 
field tensors in the momentum space. 
The tensor $T_{\alpha \beta \mu \nu}$ which should be constructed, has a 
physical dimension of the mass inversed. 

The earliest conclusion on this amplitude, the so-called Gell-Mann 
theorem~\cite{Gell-M}, states that for massless neutrinos, for both photons 
on-shell, and in the local limit of the weak interaction the amlitude 
is exactly zero. 
Really, in the center-of-mass frame the left neutrino and right 
antineutrino carry out the total momentum unity in the local limit of the 
weak interaction. 
However, two photons can not exist 
in the state with the total angular momentum equals to unity
(the Landau-Yang theorem~\cite{Lan,Yan}).
In other words, there are no covariants for constructing 
the $T_{\alpha \beta \mu \nu}$ tensor. 

With any deviation from the Gell-Mann theorem conditions, the non-zero 
amplitude arises. In the case of massive neutrino the process is 
allowed~\cite{Crew,Dode}  due to the change of the neutrino helicity. 
The electron as the lightest fermion gives the main contribution into the 
amplitude. To illustrate the Lorentz structure we present here 
the tensor $T_{\alpha \beta \mu \nu}$ 
for the case of low photon energies $(\omega \ll m_e)$
\beq
T_{\alpha \beta \mu \nu} = {{i g_A}\over {12}} \, 
{m_{\nu_i} \over m^2_e} \, \gamma_5 \,
\ee_{\alpha \beta \mu \nu}.
\label{Tmn}
\eeq

\noindent
Here $g_A$ is the axial-vector constant of the effective electron -- neutrino 
interaction in the standard model. 

In the case of non-locality of the weak interaction
the neutrino momenta become separated and the following structure 
arises~\cite{Levi,Dicu} 
\beq
T_{\alpha \beta \mu \nu} = 2 i \left (1 + {4 \over 3} \ln{m^2_W \over m^2_e}
\right ) {1 \over m^2_W} 
 \left [\gamma_\alpha \, g_{\beta \mu} (p_1 - p_2)_\nu +
\gamma_\mu \, g_{\nu \alpha} (p_1 - p_2)_\beta \right ]
 (1 - \gamma_5). 
\label{Tnl}
\eeq

\noindent
It is seen that in both cases the amplitude is suppressed, either by 
small neutrino mass in the numerator or by large $W$-boson mass in the 
denominator, and the contribution of this channel into the stellar 
energy--loss appears to be small. 

One more exotic case of non-zero amplitude is realized for off-shell 
photons~\cite{Cung,KM93}, $k_\mu f^{\mu \nu} \ne 0$, when the photon momenta 
could be included into
the tensor $T_{\alpha \beta \mu \nu}$ 
\beq
T_{\alpha \beta \mu \nu} = - {{i g_A}\over {12}} \, 
{1 \over m^2_e} \, \gamma^\rho (1 - \gamma_5) 
 \left (\ee_{\rho \alpha \mu \nu} \, k_{1 \beta}
\; 
+ \ee_{\rho \mu \alpha \beta} \, k_{2 \nu}
\right ).
\label{Trp}
\eeq

\section{
What Can the Magnetic Field Change?}

Regarding possible astrophysical applications of the process discussed, 
it should be noted that a strong 
electromagnetic field can also influence the process and could allow it. 
An external electromagnetic field opens a possibility to construct the 
amplitude because a new covariant, the electromagnetic field tensor
$F_{\mu \nu}$ arises. But the dimensionless parameter which appears in 
the amplitude is $e F_{\mu \nu}/m_e^2$. It is also the suppressing factor 
in the fields smaller than the critical Schwinger value of the electromagnetic 
field: $B_e = m^2_e/e \simeq 4.41 \cdot 10^{13}\ $ G. 
So, for the process to be significant, the fields are needed of the order or 
larger than the critical value. 
It should be mentioned that macroscopic electric 
fields above the critical value are impossible because of the $e^+ e^-$ 
pairs creation which induces the short circuit of vacuum. Only microscopic 
fields are possible in the close vicinity of heavy nuclei in the area much 
less than the elecron Compton wavelength. On the other hand, vacuum is 
stable in a magnetic field above the Schwinger value. The maximal magnetic 
field strength achieved in a laboratory is no more than $10^9$ G. 
However, it is now believed that magnetic fields stronger than the 
Schwinger value ($10^{14}, \ 10^{15}$ G or more) 
exist in astrophysical objects at the earlier stage of 
their evolution. 

Our purpose is to analyse the photon--neutrino process in a strong magnetic 
field. In the Feynman diagrams of Fig.2, describing the process, 
the double lines correspond to the 
electron propagators constructed on the base of the exact solutions of the 
Dirac equation in a magnetic field. 


\begin{minipage}[t]{100mm}

\unitlength=1.00mm
\special{em:linewidth 0.4pt}
\linethickness{0.4pt}

\vspace*{35mm}

\begin{picture}(100.00,7.00)(-20,10)
\put(30.00,38.00){\circle{13.00}}
\put(36.6,38.00){\circle*{1.3}}
\put(24.30,34.30){\circle*{1.0}}
\put(24.30,41.70){\circle*{1.0}}
\put(30.00,38.00){\circle{16.00}}

\put(34.50,38.00){\makebox(0,0)[cc]{$x$}}
\put(26.00,36.00){\makebox(0,0)[cc]{$y$}}
\put(26.00,40.50){\makebox(0,0)[cc]{$z$}}

\put(37.0,18.00){\makebox(0,0)[cc]{Fig. 2}}
\put(9.00,50.00){\makebox(0,0)[cc]{\large $\gamma$}}
\put(9.00,26.00){\makebox(0,0)[cc]{\large $\gamma$}}

\put(14.0,24.00){\photonurhalf}
\put(14.0,52.00){\photondrhalf}

\put(52.00,48.00){\makebox(0,0)[cc]{$\nu$}}
\put(52.00,28.00){\makebox(0,0)[cc]{$\bar \nu$}}

\put(37.0,38.00){\line(3,1){16.00}}
\put(37.0,38.00){\vector(3,1){10.00}}
\put(37.0,38.00){\line(3,-1){16.00}}
\put(37.0,38.00){\vector(3,-1){10.00}}

\put(76.00,38.00){\makebox(0,0)[cc]
{$+ \, (\gamma_1 \leftrightarrow \gamma_2)$}}
\end{picture}

\end{minipage}

In the recent paper~\cite{Shai} this process was studied in the case 
of relatively small magnetic field and low photon energies. The amplitude was 
obtained linearly depending on the field strength. 
An attempt of calculation in a strong field was performed earlier~\cite{Losk}
for the case of low photon energies, however, 
the amplitude presented in that paper was not gauge invariant. 

We use the effective local Lagrangian ($|q^2| \ll m^2_W$) of the 
electron -- neutrino interaction in the standard model 
\beq
{\cal L}  =  \frac{G_F}{\sqrt 2} 
\big [ \bar e \gamma_\rho (g_V - g_A \gamma_5) e \big ] \,
\big [ \bar \nu_i \gamma^\rho (1 - \gamma_5) \nu_i \big ], \; 
\label{Lef}\\
g_V =  \delta_{i e} - \frac{1}{2} + 2 sin^2 \theta_W, \;
g_A =  \delta_{i e} - \frac{1}{2}.
\nonumber
\eeq

\noindent
The process amplitude contains two essentially different parts caused by 
the vector and axial-vector parts of the electron current in the 
Lagrangian~\eq{Lef}. It should be 
mentioned that the axial-vector part of the amplitude in the local limit of 
the weak interaction contains the triangle Adler anomaly. Fortunately, the 
field-induced part of the amplitude is free of the triangle anomaly. 
It could be demonstrated by expansion of the amplitude of the process in 
Fig.2 over the external field. 


\begin{minipage}[t]{100mm}

\unitlength=1.00mm
\special{em:linewidth 0.4pt}
\linethickness{0.4pt}

\vspace*{35mm}

\begin{picture}(100.00,17.00)(-0,10)
\put(36.0,38.00){\circle*{1.0}}
\put(24.0,32.00){\circle*{1.0}}
\put(24.0,44.00){\circle*{1.0}}
\put(24.0,32.00){\line(0,1){12.00}}
\put(24.0,32.00){\line(2,1){12.00}}
\put(24.0,44.00){\line(2,-1){12.00}}

\put(57.0,13.00){\makebox(0,0)[cc]{Fig. 3}}

\put(14.0,22.00){\photonurhalf}
\put(14.0,54.00){\photondrhalf}


\put(36.0,38.00){\line(4,1){16.00}}
\put(36.0,38.00){\vector(4,1){10.00}}
\put(36.0,38.00){\line(4,-1){16.00}}
\put(36.0,38.00){\vector(4,-1){10.00}}

\put(66.00,38.00){\makebox(0,0)[cc]{$+$}}

\put(76.0,22.00){\photonurhalf}
\put(76.0,54.00){\photondrhalf}

\put(86.0,32.00){\circle*{1.0}}
\put(86.0,44.00){\circle*{1.0}}
\put(98.0,32.00){\circle*{1.0}}
\put(98.0,44.00){\circle*{1.0}}
\put(86.0,32.00){\line(0,1){12.00}}
\put(86.0,32.00){\line(1,0){12.00}}
\put(98.0,44.00){\line(0,-1){12.00}}
\put(98.0,44.00){\line(-1,0){12.00}}

\put(98.0,44.00){\line(2,1){16.00}}
\put(98.0,44.00){\vector(2,1){10.00}}
\put(98.0,44.00){\line(4,1){16.00}}
\put(98.0,44.00){\vector(4,1){10.00}}

\put(110.0,20.00){\makebox(0,0)[cc]{$+$}}

\multiput(98.00,32.00)(4.20,-4.20){3}{\line(1,-1){3.6}}
\put(98.00,32.00){\line(1,-1){2.50}}
\put(101.00,29.00){\line(1,-1){2.50}}
\put(104.00,26.00){\line(1,-1){2.50}}
\put(107.00,23.00){\line(1,-1){2.50}}

\put(120.00,38.00){\makebox(0,0)[cc]{$+ \dots$}}

\end{picture}

\end{minipage}

\noindent
Here the dashed line corresponds to the external field. 
It is obvious that only vacuum part which is zero, could contain the anomaly.

The vector part of the amplitude is interesting by itself because it defines 
the process of photon splitting in a magnetic field which is widely 
discussed in the literature.  

\section{
Details of Calculations in a Strong Magnetic Field}

The electron propagator in a magnetic field 
could be presented in the form 
\beq
S(x,y) &=& e^{\mbox{\normalsize $i \Phi (x,y)$}}\, \hat S(x-y),\label{Sxy}\\
\Phi(x,y) &=& - e \int \limits^y_x d\xi_\mu \left [ A_\mu(\xi) + \half 
F_{\mu \nu}(\xi - y)_\mu \right ],\label{Phi}
\eeq

\noindent where $e$ is the elementary charge, 
$A_\mu$ is a 4-potential of the uniform magnetic field. 
The translational invariant part $\hat S(x-y)$ of the propagator has several 
representations. It is convenient for our purposes to take it in the form of 
a partial Fourier integral expansion
\beq 
\hat S(X) &=& - \frac{i}{4 \pi} \int \limits^{\infty}_{0} 
\frac{d\tau}{th \tau}\,
\int \frac{d^2 p}{(2 \pi)^2}
\Biggl \{ 
[\prl{(p \gamma)} + m]\Pi_{-}(1 + th \tau) +
\nonumber\\
&+& [\prl{(p \gamma)} + m]\Pi_{+}(1 - th \tau) 
- \prp{(X \gamma)}\; \frac{i eB}{2\, th \tau} (1 - th^2 \tau) 
\Biggr \} 
\times 
\nonumber\\
&\times& \exp\left(- \frac{eB \zt^2}{4\, th \tau} - 
\frac{\tau(m^2 - \pp^2)}{eB} - i \prl{(pX)} \right), 
\label{Sx}\\[3mm]
d^2 p &=& dp_0 dp_3, \quad \Pi_\pm = \half (1 \pm i \gamma_1 \gamma_2),
\quad \Pi^2_\pm = \Pi_\pm, \quad [\Pi_\pm, \prl{(a \gamma)}] = 0,
\nonumber
\eeq
 
\noindent where $\gamma_\alpha$ are the Dirac matrices in the standard 
representation, the four-vectors with the indices $\bot$ and $\parallel$ 
belong 
to the \{1, 2\} plane and the Minkowski \{0, 3\} plane correspondingly, 
when the field $\bf B$ is directed along the third axis. Then for arbitrary 
4-vectors $a_\mu$, $b_\mu$ one has
\beq
\prp{a} &=& (0, a_1, a_2, 0), \quad  \prl{a} = (a_0, 0, 0, a_3), \nonumber \\
\prp{(ab)} &=& (a \ff b) =  a_1 b_1 + a_2 b_2 , \quad 
\prl{(ab)} = (a \tff b) = a_0 b_0 - a_3 b_3, 
\label{ab}
\eeq

\noindent where the matrices are introduced
$\ff_{\alpha \beta} = (\varphi \varphi)_{\alpha \beta}$,\,  
$\tff_{\alpha \beta} = 
(\tilde \varphi \tilde \varphi)_{\alpha \beta}$, connected by the relation 
$\tff_{\alpha \beta} - \ff_{\alpha \beta} = 
g_{\alpha \beta} = diag(1, -1, -1, -1)$, 
$\varphi_{\alpha \beta} =  F_{\alpha \beta} /B,\; 
{\tilde \varphi}_{\alpha \beta} = \frac{1}{2} \varepsilon_{\alpha \beta
\mu \nu} \varphi_{\mu \nu}$ are the dimensionless tensor of the external 
magnetic field and the dual tensor, 
$(a \ff b) = a_\alpha \ff_{\alpha \beta} b_\beta$.

In spite of the translational and gauge noninvariance of the phase 
$\Phi(x, y)$ in the propagator~\eq{Sxy}, the total phase of three propagators 
in the loop is translational and gauge invariant
\beq 
\Phi(x, y) + \Phi(y, z) + \Phi(z, x) = - \frac{e}{2} (z - x)_\mu 
F_{\mu \nu} (x - y)_\nu. \nonumber
\eeq

\noindent Some papers were known where an incorrect operating with these 
phases was a source of mistakes.

The amplitude of the photon -- neutrino process $\ggnunu$ takes the form
\beq
\M &=& e^2 {G_F \over \sqrt{2}} \int d^4 X\, d^4 Y\, Sp \{\hat j 
\hat S(Y) \hat \ee'' 
\hat S(-X-Y) \hat  \ee' \hat S(X)\} \times \nonumber \\
&\times& e^{- i e\,(X F Y)/2}\; e^{i (k' X - k'' Y)} + 
(\ee', k' \leftrightarrow \ee'', k''),
\label{M}
\eeq

\noindent where $j$ is the neutrino current in the momentum space, 
$\ee',\; k'$ and $\ee'',\;k''$ are the polarization vectors and
the 4-momenta of initial photons, 
$X = z - x, \, Y = x - y$. 

Manipulations with the propagator~\eq{Sx} in the three-point loop leads to 
a very cumbersome expression in a general case. Relatively simple results 
were obtained for the process of photon splitting 
only in the two limits of a weak field~\cite{Ad71} and of the 
strong field with collinear kinematics~\cite{Baier}. 

It is advantageous to use the asymptotic expression of the electron 
propagator for an analysis of the process amplitude in the strong field. 
This asymptotic could be easily derived 
from Eq.~\eq{Sx} by evaluation of the integral over $\tau$ in the limit

\noindent 
$eB /\vert m^2 - \prl{p}^2 \vert \gg 1$. In this case the propagator 
takes the simple form
\beq
\hat S(X) \simeq S_a(X) = \frac{i eB}{2 \pi} \exp (- \frac{eB \zt^2}{4}) 
\int \frac{d^2 p}{(2 \pi)^2}\, \frac{\prl{(p\gamma)} + m}{\prl{p}^2 - m^2}
\Pi_{-}e^{-i \prl{(pX)}}, 
\label{Sa}
\eeq

\noindent 
which was obtained for the first time in Ref.~\cite{Skob}. 

Substituting the propagator~\eq{Sa} into the amplitude one obtains that two 
parts of it which differ by the photon interchange 
are linear on the field $B$, but they cancel each other. 
Thus, the asymptotic form 
of the electron propagator~\eq{Sa} only shows that the 
linear-on-field part of the amplitude is zero and provides no way of 
extracting the next term of expansion over the field strength. 

We have also studied the process $\gamma \gamma\to \nu \bar\nu$ with the 
asymptotic propagator~\eq{Sa} for more 
general case when the effective two-fermion -- two-neutrino interaction 
contains 
the scalar, pseudoscalar, vector, and axial-vector connections. 
We have obtained that only the scalar connection 
gives the linear growth with $B$ 
in the amplitude. It is known that the effective scalar 
two-fermion -- two-neutrino interaction 
arises in the models with supersymmetry, left-right symmetry, 
and with leptoquarks (a fermion in the loop is a quark in the last case). 
Thus, the process $\gamma \gamma\to \nu \bar\nu$ could be a channel 
of creation of sterile neutrinos, amplified by a strong magnetic field.  
It could be an interesting direction for further investigations. 

As the analysis shows, in the standard model the linear increase of the 
amplitude with the field takes place in the next order of perturbation theory, 
in the process $\gamma \gamma\to \nu \bar\nu \gamma$. 
The probability of this process contains 
the extra factor $\alpha (B/B_e)^2$ as compared with the process 
$\gamma \gamma\to \nu \bar\nu$. Thus, for the field strength 
$B > 10^{15}\, $ G, the process with additional photon would dominate. 

\section{
Partial Amplitudes of the Process
$\gamma \gamma\to \nu \bar\nu $} 

The next term of expansion of the amplitude over the field strength
can be found by the insertion of two 
asymptotic~\eq{Sa} and one exact propagator~\eq{Sx} into the amplitude~\eq{M}, 
with all interchanges. 
It is worthwhile to turn from the general 
amplitude~\eq{M} to the partial polarization amplitudes. 
For this purpose we use the orthogonal basis of 4-vectors
\beq 
b_\alpha^{(1)} = \frac{(k \varphi)_\alpha}
{\sqrt{k_{\mprp}^2}}, 
\;
b_\alpha^{(2)} = \frac{(k \tilde \varphi)_\alpha}
{\sqrt{k_{\mprl}^2}}, 
\;
b_\alpha^{(3)} = \frac{k^2 (k \ff)_\alpha - (k \ff k) k_\alpha}
{\sqrt{k^2 k_{\mprp}^2 k^2_{\mprl}}}, 
\; 
 b_\alpha^{(4)} = \frac{k_\alpha}{\sqrt{k^2}}, 
\label{b's}
\eeq

\noindent 
with the definitions of Eq.~\eq{ab}.
The vectors $b_\alpha^{(1)}$ and $b_\alpha^{(2)}$ 
correspond to the stationary photon states with definite dispersion 
relations in a magnetic field, and an arbitrary polarization vector 
$\ee_\alpha$ of a photon with the momentum $k$ can be expanded over these 
two vectors. 
The neutrino current $j$ in the case of massless neutrinos is orthogonal to 
the total 4-momentum of the neutrino -- antineutrino pair, and it could be 
expanded over the three vectors, 
$b_\alpha^{(1)}\;$, $b_\alpha^{(2)}$ and $b_\alpha^{(3)}$, where $k$ is 
the total 4-momentum of the pair. 

Making these expansions in the amplitude~\eq{M}, one obtains 
18 independent partial amplitudes
\beq 
\M = \sum_{\scriptstyle\lm = 1,2,3 \atop\scriptstyle\lm',\lm'' = 1,2} 
(b^{(\lm)} j) (b^{(\lm')} \ee') (b^{(\lm'')} \ee'')
\; \left (g_V \M^V_{\lm\lm'\lm''} + g_A \M^A_{\lm\lm'\lm''}\right )
\label{Mva}
\eeq

\noindent 
There are 6 amplitudes $\M^V$ for $\lm = 1,2$ which
depict the process of photon splitting, with the substitution 
$G_F/\sqrt{2} \to e$. 

We have obtained the following expressions for the partial amplitudes, 
to the terms of order $1/B$ 
\beq 
\M^V_{111} &=& \M^A_{111} = \M^A_{211} = \M^V_{311} = 0,
\label{MV111} \\
[4mm]
\M^V_{112} &=& 
i \;\frac{2 \alpha}{\pi} \;\frac{G_F}{\sqrt{2}} \;
\frac{(k' \varphi k'')(k' \tilde \varphi k'')}
{[
\prl{(k')^2}\prp{(k'')^2}\prp{k^2}
]^{1/2}}\; \HHi, 
\label{MV112} \\
[4mm]
\M^A_{112} &=& 
i \;\frac{2 \alpha}{\pi} \;\frac{G_F}{\sqrt{2}} \;
\frac{(k' \varphi k'')(k' \tff k'')}
{[
\prl{(k')^2}\prp{(k'')^2}\prp{k^2}
]^{1/2}}\; \HHi, 
\label{MA112} \\
[4mm]
\M^V_{122} &=& 
i \;\frac{2 \alpha}{\pi} \;\frac{G_F}{\sqrt{2}} \;
\frac{(k' \tff k'')}
{[\prl{(k')^2}\prl{(k'')^2}\prp{k^2}]^{1/2}}
\Biggl \{ (k \ff k'') \HHi 
+ (k \ff k') \HHii \Biggr \}, 
\label{MV122} \\
[4mm]
\M^A_{122} &=& 
i \;\frac{2 \alpha}{\pi} \;\frac{G_F}{\sqrt{2}} \;
\frac{(k' \tilde \varphi k'')}
{[\prl{(k')^2}\prl{(k'')^2}\prp{k^2}]^{1/2}}
\Biggl \{ (k \ff k'') \HHi 
- (k \ff k') \HHii \Biggr \}. 
\label{MA122} 
\eeq

\noindent 
Here $k',\;k''$ are the photon momenta, $k$ is the total momentum of the 
$\nu \bar \nu$ pair, and the function is introduced 
\beq
H(z) &=&\frac{z}{\sqrt{z - 1}} \arctan \frac{1}{\sqrt{z - 1}} - 1, \quad z > 1,
\nonumber \\
H(z) &=& - \half \left ( \frac{z}{\sqrt{1-z}}
\ln \frac{1 + \sqrt{1-z}}{1 - \sqrt{1-z}} + 2 + 
i \pi \frac{z}{\sqrt{1-z}} \right ), \quad z < 1.
\label{Hz}
\eeq

\noindent 
The remaining amplitudes have a more complicated form and will be published 
elsewhere. 

\section{
Photon Splitting in a Strong Magnetic Field} 

As was mentioned above, 6 partial amplitudes of the process $\ggnunu$ 
describe also the photon splitting $\ggg$. This process has its own long 
history of investigations, see e.g.~\cite{PapRit89}. 
Recent progress in astrophysics 
has drawn attention again to the photon splitting induced by a magnetic 
field~\cite{Hard}. 
The study of this process in a strong 
magnetic field~\cite{Baier,Ad96,W} has so far considered the 
collinear limit, when the only allowed 
transition with respect to photon polarizations is, $\alw$ (in Adler's 
notation~\cite{Ad71}). However, photon dispersion in a strong 
magnetic field, $B \gg B_e$, leads to significant deviations 
from the collinearity of the kinematics of this process. 
This is due to the fact that the eigenvalues of the 
photon polarization operator (the photon effective mass squared) 
become large near the so-called cyclotron 
resonances~\cite{Shab}. 
On the other hand, a large value of the polarization operator near the 
resonance requires taking account of large radiative 
corrections which reduce to a renormalization of the photon 
wave-function  
\beq
\ee_{\alpha}^{(\lm)} \to \ee_{\alpha}^{(\lm)} \sqrt{Z_\lm}, \quad 
Z^{-1}_\lm = 1 - \frac{\partial \Pl}{\partial \omega^2}, \quad 
\lm = \parallel, \perp. 
\label{Z}
\eeq    

\noindent Here 
$\ee_\alpha^{(\mprl)} = b^{(1)}_\alpha,\;
\ee_\alpha^{(\mprp)} = b^{(2)}_\alpha$
are the polarization 
four-vectors of the photon modes and $\Pl$ are the eigenvalues of the 
photon polarization operator, corresponding to these modes~\cite{Shab}. 

Both the effect of noncollinearity and 
radiative corrections have not, so far, been taken into account. 
Substituting $G_F/\sqrt{2} \to e$ in the amplitudes~\eq{MV112} 
and~\eq{MV122} we obtain from $\M^V_{112}$ the amplitude of the 
process $\frb$, forbidden in the collinear limit and from $\M^V_{122}$ the 
amplitude of the allowed process $\alw$.
As for remaining amplitudes, we note that $\M (\zer)$ 
is equal to zero in this approximation, see Eq.~\eq{MV111}. 
On the other hand, the photon of the 
$\perp$ mode due to its dispersion can split into two photons only in the 
kinematic region $\prl{k^2} > 4 m^2$ where the tree-channel 
$\gamma_{\mprp} \to e^+ \,e^-$~\cite{Klep} strongly dominates.

Thus we will analyse further the photon splitting of the $\|$ mode 
in the region $\prl{k^2} < (m + \sqrt{m^2 + 2 e B})^2$ where the tree-channel 
$\gamma_{\mprl} \to e^+ \,e^-$ does not exist. 
In the formal limit of the collinearity of the photon momenta, the amplitude 
$\M (\frb)$ goes to zero while the amplitude 
$\M (\alw)$ coincides with the amplitude obtained in 
Ref.~\cite{Baier}. 
However, the collinear limit $\vert q^2 \vert/\omega^2\ll 1$ is inadequate in 
the strong field
${\vert q^2 \vert}/{\omega^2} \simeq ({\alpha}/{3 \pi})({B}/{B_e}) 
\sim 1$, when ${B}/{B_e} \simeq {3 \pi}/{\alpha}\sim 10^3$.

Although the process involves three particles, its amplitude is not a 
constant, 
because it contains the external field tensor in addition to the photon 
4-momenta. The general expression for the splitting probability can be 
written in the form
\beq
W (\gamma_\lm \to \gamma_{\lm'} \gamma_{\lm''}) &=& 
\frac{g}{32 \pi^2 \omega} \int \vert \M
 (\gamma_\lm \to \gamma_{\lm'} \gamma_{\lm''}) 
\vert^2
Z_\lm Z_{\lm'} Z_{\lm''} \times
\nonumber\\
&\times& \; \delta(\omega_\lm({\bf k}) - \omega_{\lm'}({\bf k'}) - 
\omega_{\lm''}({\bf k} - {\bf k'})) 
\frac{d^3 k'}{\omega_{\lm'} \omega_{\lm''}},
\label{defW}
\eeq

\noindent where the factor $g = 1 - {1 \over 2}\delta_{\lm' \lm''}$ is 
inserted to account for possible identity of the final photons. 
The integration over phase space of two final photons in Eq.~\eq{defW} 
has to be performed using the photon energy dependence on the 
momenta, $\omega = \omega_\lm({\bf k})$, which can be found from 
the dispersion equations
\beq
\omega_\lm^2({\bf k}) - {\bf k}^2 - {\cal P}^\lm = 0.
\label{disp}
\eeq

A calculation of the splitting probability~\eq{defW} is rather complicated 
in the general case.  
In the limit $m^2 \ll \omega^2 \sin^2 \theta \ll eB$,
where $\theta$ is an angle between the initial photon momentum
$\bf k$ and the magnetic field direction, we derive the 
following analytical expression for the probability of the channel $\frb$:
\beq
W &\simeq& \frac{\alpha^3 \omega \sin^2 \theta}{16} (1 - x)
[1 - x + 2 x^2 + 2(1 - x)(1 + x)^2 \ln (1 + x) -
\nonumber\\[2mm]
&-& 2 x^2 \,\frac{2-x^2}{1-x} \, \ln \frac{1}{x}], \qquad
x = \frac{2 m}{\omega \sin \theta} \ll 1. 
\label{Wa}
\eeq

Within the same approximation we obtain the spectrum of 
final photons in the frame where the initial photon momentum is orthogonal 
to the field direction: 
\beq
\frac{d W}{d \omega'} \simeq \frac{\alpha^3}{2}\cdot
\frac{\sqrt{(\omega - \omega')^2 - 4 m^2}}
{\omega' + \sqrt{(\omega' - \omega)^2 - 4 m^2}}, 
\label{spectr}\\
\frac{\omega}{2} - \frac{2 m^2}{\omega} < \omega' < \omega - 2 m, 
\nonumber
\eeq

\noindent where $\omega, \omega'$ are the energies of the initial and final 
photons of the $\parallel$ mode. 

We have made numerical calculations of the process probabilities 
below and near the pair-creation threshold 
for both channels~\cite{CKM}, which are valid in the limit
$\omega^2 \sin^2 \theta \ll eB$. 
In this region the channel $\alw$ (allowed in the 
collinear limit) is shown to dominate the channel $\frb$ (forbidden in 
this limit). The probability 
obtained without considering the noncollinearity of the kinematics and 
radiative corrections is shown to be inadequate. 
For example, this probability becomes infinite just above the threshold. 
Both channels give essential contributions 
to the probability at high photon energies, with the 
``forbidden'' channel dominating. 
It should be stressed that taking account of the photon polarization 
leads to the essential dependence of the splitting probabilities on the 
magnetic field, while the amplitudes do not depend 
on the field strength value. 

\section{Conclusions}

In this paper we have considered the photon-neutrino process 
$\gamma \gamma\to \nu \bar\nu $ and photon splitting 
in a strong magnetic field. 
It is shown that various types of the neutrino -
electron effective interactions lead to different dependences of the 
amplitudes on the field strength.
The partial polarization amplitudes are calculated within the standard model 
in the limit of a strong field. The amplitudes do not depend on the 
field strength in this limit.

Using the vector parts of the amplitudes, the photon 
splitting $\ggg$ is investigated.
The collinear limit is shown to be an inadequate approximation for 
this process
in a strong magnetic field ($B \gg B_e$), because of the significant 
deviation of 
the photon dispersion in the strong field from the vacuum dispersion.
The ``allowed'' channel $\alw$ is not comprehensive for the splitting in 
the strong field. The ``forbidden'' channel $\frb$ is also essential, moreover,
it dominates at high energies of the initial photon. 
The photon splitting probabilities are calculated 
in the strong field limit for both channels. The probabilities depend 
essentially on the field strength value, 
due to the photon polarization in the strong magnetic field.

\vspace{10mm} 

\section{Acknowledgements}

The authors are grateful to V.A.~Rubakov for useful discussions of a 
problem of large radiative corrections in the vicinity of a cyclotron 
resonance.
A.K. expresses his deep gratitude to the organizers of the 
Ringberg Neutrino Euroconference for the possibility to participate in it.
This work was supported in part by the INTAS Grant N~96-0659  
and by the Russian Foundation for Basic Research Grant N~98-02-16694.
The work of A.K. and N.M. was supported in part by the
International Soros Science Education Program under the Grants 
N~d98-179 and N~d98-181.

\section{References}

\end{document}